\documentclass[
  twocolumn,
  nofootinbib,
 amsmath,amssymb,
 aps, 
 groupedaddress,
 floatfix,
]{revtex4-1}
\usepackage{setspace}
\usepackage{lipsum}
\usepackage{color}
\usepackage{booktabs}
\usepackage{verbatim}
\usepackage{soul}
\usepackage{afterpage}
\usepackage{multirow,dcolumn}
\usepackage[hidelinks]{hyperref}
\hypersetup{
colorlinks= true,
menucolor = {black},
linkcolor = {black},
citecolor= {blue},
urlcolor = {blue}
}
\usepackage{cellspace}
\usepackage{cleveref}
\usepackage{tabularx}
\usepackage{tabu}
\usepackage{graphicx}
\usepackage{bm}
\binoppenalty=\maxdimen
\relpenalty=\maxdimen

\usepackage{xfrac}    
\usepackage{nicefrac} 
\usepackage{xcolor}

\newcommand{\Xstate}[1]{\ensuremath{X~^1\Sigma_g^+#1}}
\newcommand{\bstate}[1]{\ensuremath{b~^3\Sigma_u^+#1}}

\newcommand{\bra}[1]{\langle#1|}
\newcommand{\ket}[1]{|#1\rangle}

\newcommand{\abs}[1]{\left|#1\right|}
\newcommand{\sqbr}[1]{\left[#1\right]}
\newcommand{\dee}{\, \ensuremath{\mathrm{d}}}

\newcommand{\dd}[2]{\frac{\dee#1}{\dee#2}} 

\newcommand{\minus}{\text{\scalebox{0.75}[1.0]{$-$}}}
\usepackage{titlesec}

\begin{document}

\title{Isotopic and vibrational-level dependence of H$_{\text{\textbf 2}}$ dissociation by electron impact} 
\author{Liam H. Scarlett$^1$}
\email{liam.scarlett@postgrad.curtin.edu.au}
\author{Dmitry V. Fursa$^1$}
\author{Jack Knol$^1$}
\author{Mark C. Zammit$^2$}
\author{Igor Bray$^1$}

\affiliation{$^1$Curtin Institute for Computation and Department of Physics and Astronomy, Curtin University, 
Perth, Western Australia 6102, Australia}
\affiliation{$^2$Theoretical Division, Los Alamos National Laboratory, Los Alamos, New Mexico 87545, USA}
\date{\today}

\begin{abstract}
  The low-energy electron-impact dissociation of molecular hydrogen has been a source of disagreement between 
  various calculations and measurements for decades. 
  Excitation of the ground state of H$_2$ into the dissociative \bstate{} state is now well understood,
  with the most recent measurements being in excellent agreement with the molecular convergent close-coupling
  (MCCC) calculations of both integral and differential cross sections
  (2018~\textit{Phys.~Rev.~A}~\textbf{98}~{062704}). However, in the absence of similar measurements for 
  vibrationally-excited or isotopically-substituted H$_2$, cross sections for dissociation of these
  species must be determined by theory alone.
  We have identified large discrepancies between MCCC calculations and the recommended $R$-matrix cross sections
  for dissociation of vibrationally-excited H$_2$, D$_2$, T$_2$, HD, HT, and DT 
  (2002~\textit{Plasma~Phys.~Contr.~F.}~\textbf{44}~{1263--1276,2217--2230}), with disagreement in both the isotope
  effect and dependence on initial vibrational level.
  Here we investigate the source of the discrepancies, and discuss the consequences for plasma models 
  which have incorporated the previously recommended data.
\end{abstract}

\maketitle

In low temperature plasmas, electron-impact dissociation of molecular hydrogen into neutral fragments proceeds 
almost exclusively via excitation of the dissociative \bstate{} state:
\begin{equation}\label{eq:react}
  \text{e}^\minus + \text{H}_2(\Xstate{},v) \to \text{H}_2(\bstate{}) + \text{e}^\minus \to 2\text{H} + \text{e}^\minus.
\end{equation}
Cross sections and rate coefficients for this process are needed to accurately model astrophysical, scientific, 
and technological plasmas where hydrogen is present in its molecular form~\cite{Vorobyov2020,Shin2014,Broks2005}.
The importance of the reaction~(\ref{eq:react}) and the relative simplicity it presents to theory and experiment 
has led to it being one of the most studied processes in electron-molecule scattering.
Despite this, for decades there was no clear agreement between any theoretically- or experimentally-determined
cross sections. The largest molecular convergent close-coupling (MCCC) calculations~\cite{ZSFB17} for scattering on the
ground vibrational level ($v=0$) of H$_2$ were up to a factor of two smaller than the recommended cross sections, but
the situation was resolved when newer measurements were found to be in near-perfect agreement with the 
MCCC results~\cite{TOF18,TOF18Long}. Recent $R$-matrix calculations have also confirmed the MCCC results for scattering on 
the $v=0$ level of H$_2$~\cite{Meltzer2020}.

It is also important to determine accurate cross sections for dissociation of vibrationally-excited and 
isotopically-substituted hydrogen molecules, due to their presence in fusion and
astrophysical plasmas. For these species, however, the absence of experimental data means it is up to 
theory to make recommendations alone.
The primary distinguishing factor between the numerous calculations of the \bstate{}-state
cross section~%
\cite{FM80,SC85,Baluja1985a,Lima1984,Lima1985,RS88,BTM90,ST98,CJLCWA01,TT01,CPL05,GT05,ZFB14,Scarlett17,Scarlett17b3,Tapley18DE,LaricchiutaCJ04}
is the treatment of the electronic dynamics. There is generally a consensus in the literature that the nuclear dynamics
of the dissociative transition can be treated using the standard adiabatic-nuclei (AN) approximation 
(as reviewed by \citet{Lane80}). We have taken this approach previously to study the dissociation of
vibrationally-excited H$_2^+$ and its isotopologues, finding good agreement with measurements of both 
integral and energy-differential (kinetic-energy-release) cross sections~\cite{ZFB14,Scarlett17}.

The cross sections for dissociation of vibrationally-excited H$_2$, HD, and D$_2$ recommended in the well-known
reviews of \citet{Yoon2010,Yoon2014} come from the $R$-matrix calculations of \citet{TT02,TT02-mixed}
(hereafter referred to collectively as TT02). 
The results of TT02, which also include HT, DT, and T$_2$, have long been considered the most accurate dissociation 
cross sections for H$_2$ and its isotoplogues, and are widely used in applications.
During our recent efforts to compile a comprehensive set of vibrationally-resolved cross sections for electrons 
scattering on vibrationally-excited and isotopically-substituted H$_2$~\cite{Scarlett20adndt1,Scarlett20adndt2},
it has become apparent that there are major discrepancies between the MCCC calculations and the $R$-matrix
calculations of TT02. Interestingly, the two methods are similar in their treatment of the electronic dynamics and
differ primarily in their treatment of the nuclear dynamics, leading to conflicting isotopic and 
vibrational-level dependencies in the calculated cross sections. The formalism applied by TT02 was previously
developed by \citet{TT01} (TT01).

Here we compare the standard AN method adopted in the MCCC calculations
with the alternative formulation suggested by TT01, and determine the origin of the disagreement between the two 
approaches. We will argue  in favour of
the MCCC results, which will be of some interest to those who have previously recommended the $R$-matrix 
data~\cite{Yoon2008,Yoon2010,Yoon2014,Raju2006}
or implemented it in their 
models~\cite{Ziegler2018,Vorobyov2020,Sormani2018,Smith2017,Schleicher2008,Richings2014,Micic2012,Micic2013,
Matsukoba2019,MacKey2019,Glover2009,Glover2008,Coppola2013,Clark2019,Clark2011,Broks2005,Shin2014,Moravej2004}.
We limit the discussion here to dissociation through the \bstate{} state, but note that we have 
previously~\cite{Scarlett19diss} performed more detailed dissociation calculations for vibrationally-excited H$_2$ 
including all important pathways to dissociation into neutral fragments from low to high incident energies. 
These calculations can be readily extended to include the isotopologues in the future.

We first describe the standard AN treatment of dissociation, which is a straightforward adaption of the method for
non-dissociative excitations described by \citet{Lane80}.
We use SI units throughout for consistent comparison with TT01's formulas.
In terms of the electronic partial-wave $T$-matrix elements defined by \citet{Lane80},
the expression for the energy-differential cross section for dissociation of the vibrational level $v$
into atomic fragments of asymptotic kinetic energy $E_\text{k}$ in the standard AN method is
\begin{equation}\label{eq:cs-ST}
\dd{\sigma}{E_\text{out}} = \frac{\pi}{k_\text{in}^2}\sum_{\substack{\ell'm'\\\ell m}}
\abs{\bra{\nu_{E_\text{k}}}T_{\ell'm',\ell m}(R;E_\text{in})\ket{\nu_v}}^2,
\end{equation}
where $E_\text{in}$ and $k_\text{in}$ are the incident projectile energy and wavenumber, 
and $\nu$ are the vibrational wave functions.
The energies of the scattered electron and dissociating fragments are related by
\begin{equation}
  E_\text{k} = E_\text{in} - D_v - E_\text{out},
\end{equation}
where $D_v$ is the threshold dissociation energy of the vibrational level $v$~\cite{ST98}.
This relationship makes it possible to treat the energy-differential cross section as a function
of either $E_\text{out}$ or $E_\text{k}$.
When this method has been applied in previous work~%
\cite{FM80,SC85,Baluja1985a,Lima1984,Lima1985,RS88,BTM90,ST98,CJLCWA01,CPL05,GT05,ZFB14,Scarlett17,Scarlett17b3,Tapley18DE,LaricchiutaCJ04},
Eq.~(\ref{eq:cs-ST}) is not derived explicitly for the case of dissociation since the derivation follows
exactly the same steps summarized by \citet{Lane80} for the non-dissociative vibrational-excitation cross section.
The only difference is the replacement of the final bound vibrational wave function with an appropriately-normalized
continuum wave function $\nu_{E_\text{k}}(R)$. In principle the continuum normalization is arbitrary so long 
as the density of final states is properly accounted for. 
According to standard definitions~\cite{ClaudeCohenQM}, the density of states $\rho$ for the vibrational continuum
satisfies the following relation:
\begin{equation}\label{eq:unity-E}
  \int \nu_{E_\text{k}}(R)\nu_{E_\text{k}}(R')\rho(E_\text{k})\dee E_\text{k} = \delta(R-R'),
\end{equation}
giving a clear relationship between the continuum-wave normalization and density.
Since the formulas presented by \citet{Lane80} for non-dissociative excitations are written in terms of bound vibrational
wave functions, with a density of states equal to unity (by definition), the most straightforward adaption to dissociation
simply replaces them with continuum wave functions normalized to have unit density as well.
Indeed, many of the previous works~\cite{RS88,ST98,Celiberto1989,Celiberto1994,LaricchiutaCJ04,Scarlett17}
which have applied the AN (rather than FN) method to dissociation explicitly state
that the continuum wave functions are energy normalized, which implies unit density and the following resolution of unity:
\begin{align}
  \int_0^\infty \nu_{E_\text{k}}^*(R)\nu_{E_\text{k}}(R')\dee E_\text{k} &= \delta(R-R').  \label{eq:norm-en-E}
\end{align}
The works which have applied the FN method also implicitly assume energy normalization, since the FN approximation
utilizes Eq.~(\ref{eq:norm-en-E}) to integrate over the dissociative states analytically.
Note that Eq.~(\ref{eq:norm-en-E}) implies
the functions $\nu_{E_\text{k}}$ have dimensions of
$1/\!\sqrt{\text{energy}\cdot\text{length}}$.
The bound vibrational wave functions are normalized according to
\begin{equation}
  \int_0^\infty\nu_v^*(R)\nu_{v'}(R)\dee R = \delta_{v'v},
\end{equation}
and hence they have dimensions of $1/\!\sqrt{\text{length}}$.
The electronic $T$-matrix elements defined by \citet{Lane80} are dimensionless,
and the integration over $R$ implied by the bra-kets in Eq.~(\ref{eq:cs-ST}) cancels the combined
dimension of $1/\text{length}$ from the vibrational wave functions,
so it is evident that the right-hand side of
Eq.~(\ref{eq:cs-ST}) has dimensions of $\text{area}/\text{energy}$
as required (note that $1/k_\text{in}^2$ has dimensions of area).

The standard AN/FN approaches to calculating dissociation cross sections have been applied extensively in the 
literature~\cite{FM80,SC85,Baluja1985a,Lima1984,Lima1985,RS88,BTM90,ST98,CJLCWA01,TT01,CPL05,GT05,ZFB14,Scarlett17,Scarlett17b3,Tapley18DE,LaricchiutaCJ04}.
They are also consistent with well-established methods for computing bound-continuum radiative lifetimes 
or photodissociation cross sections, which replace discrete final states with dissociative vibrational wave functions.
The latter are either energy normalized~\cite{ZSCFKBFH717}, or normalized to unit asymptotic amplitude with the
energy-normalization factor included explicitly in the dipole matrix-element formulas~\cite{Dalgarno1970,Fantz2000a}.

TT01 criticized the standard technique, claiming that a proper theoretical formulation for dissociation did not
exist, and suggested that a more rigorous derivation for the specific case where there are three fragments in the
exit channels is required. We have performed our own derivation following the ideas laid out by TT01 and found that
they lead directly to Eq.~(\ref{eq:cs-ST}). 
However, TT01 arrived at an expression which is markedly different:
\begin{equation}\label{eq:cs-TT}
  \dd{\sigma}{E_\text{out}} = \frac{m_\text{H}}{4\pi^3 m_\text{e}}\frac{E_\text{k}}{E_\text{in}}
  \sum_{\substack{\ell'm'\\\ell m}}\abs{\bra{\nu_{E_\text{k}}}T_{\ell'm',\ell m}(R;E_\text{in})\ket{\nu_v}}^2.
\end{equation}
Here $m_\text{H}$ is the hydrogen nuclear mass, which is replaced with the deuteron or triton
mass in their later investigation into dissociation of D$_2$ and T$_2$~\cite{TT02}.
Comparing Eqs.~(\ref{eq:cs-ST}) and (\ref{eq:cs-TT}), we see that TT01's formula is different
by a factor of $m_\text{H}E_\text{k}/2\pi^4\hbar^2$ (the $T$-matrix elements here are the same as those 
in Eq.~(\ref{eq:cs-ST})). 
The distinguishing feature of TT01's approach and the reason for the mass-dependence in their formula was said to be
the explicit consideration of the density of dissociating states.
We have two major concerns here: firstly, the energy-normalized wave functions 
used in Eqs.~(\ref{eq:cs-ST}) and (\ref{eq:cs-TT}) have unit density so it is unusual that taking this into account
should have any effect, and secondly, TT01's expression for the
energy-differential cross section has dimensions of
\begin{equation}
  \dim\sqbr{\dd{\sigma}{E_\text{out}}} = \frac{\text{1}}{\text{energy}},
\end{equation}
which suggests an error in the derivation.
As a result, the integral cross sections for scattering on H$_2(v=0)$ presented in TT01 and for
vibrationally-excited H$_2$, HD, D$_2$, HT, DT, and T$_2$ presented in TT02 using the same method appear to 
be  incorrect.

TT01's derivation uses a density of states corresponding to (three-dimensional) momentum normalization.
However, rather than calculating momentum-normalized vibrational wave functions, TT01 use energy-normalized 
functions and apply a correction factor 
\begin{equation}\label{eq:chi}
  \xi^2 = 2\hbar\left(\frac{E_\text{k}}{m_\text{H}}\right)^{1/2} = \frac{\hbar p_\text{k}}{\mu}
\end{equation}
to the cross section to account for a conversion from energy to momentum normalization. 
Since the vibrational wave functions are one-dimensional, it is not obvious how to normalize them
to three-dimensional momentum.
Although TT01 do not state
explicitly how they choose to define the momentum normalization, 
Eq.~(\ref{eq:chi}) corresponds to a conversion from energy 
normalization to (one-dimensional) wavenumber normalization~\cite{Morrison1994}.
The density of states for this choice of normalization is
\begin{equation}
  \rho(E_\text{k}) = \frac{\mu}{\hbar p_\text{k}},
\end{equation}
which cancels exactly with the correction factor~(\ref{eq:chi}).
This is to be expected since Eq.~(\ref{eq:unity-E})
shows clearly that any factors applied to the continuum wave functions to change the normalization must lead to the inverse
factor (squared) being applied to the density of states.
It is the mismatch between continuum normalization and density of states
which leads to some of the additional factors, such as the molecular mass, 
in TT01's final cross section formula. Using a consistent normalization and density of states it is possible
to follow the remaining steps taken by TT01 in their derivation and arrive at an expression identical to the
standard formula~(\ref{eq:cs-ST}). We have provided our own derivation in the supplementary materials. 

The novelty of the reformulated $R$-matrix approach has been acknowledged numerous times and the
results have been widely adopted.
Perhaps in part due to being recommended by \citet{Yoon2008,Yoon2010} and included in the 
Quantemol database~\cite{Tennyson2017}, the cross sections and rate coefficients 
given by TT02 have been applied in a number of different plasma models, most notably in the astrophysics 
community~\cite{Ziegler2018,Vorobyov2020,Sormani2018,Smith2017,Schleicher2008,Richings2014,Micic2012,Micic2013,Matsukoba2019,MacKey2019,Glover2009,Glover2008,Coppola2013,Clark2019,Clark2011,Broks2005,Shin2014,Moravej2004}.
The formalism of TT01 was also used by \citet{Gorfinkiel2002} to study the
electron-impact dissociation of H$_2$O, and it has been reiterated a number of times
that this method is necessary to accurately treat dissociation in the AN 
approximation~\cite{Tennyson2002,Burke2005a,Chakrabarti2007}.
\citet{Gorfinkiel2002} found in particular that for some dissociative transitions in H$_2$O the formalism of TT01
gives results up to a factor of two different to the FN method even 10~eV above threshold. If correct this result
would invalidate the use of the FN method in dissociation calculations, for example in the $R$-matrix
calculations of Refs.~\cite{BTP14,Chakrabarti2009,Chakrabarti2017,BTP14}.

\begin{figure}[t!]
  \centering
  \includegraphics{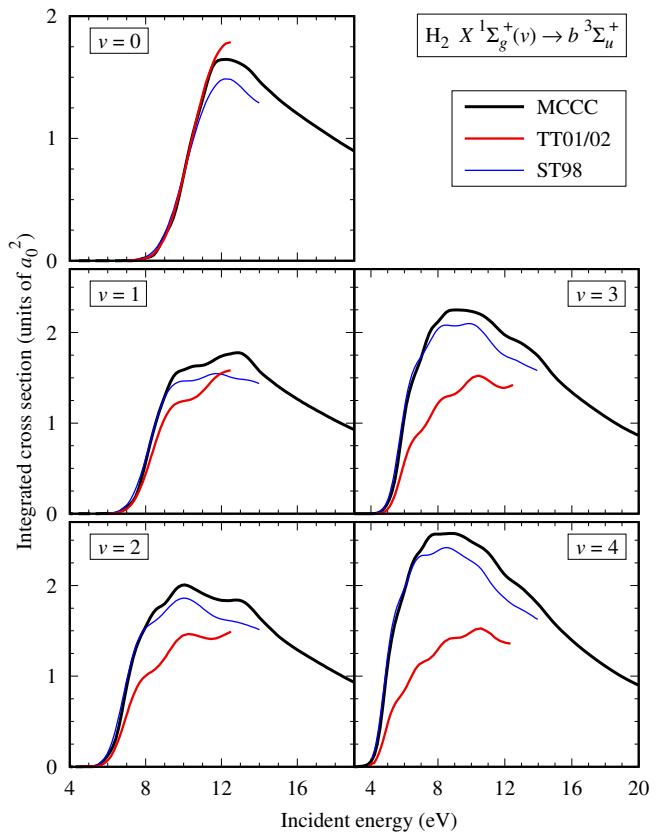}
  \caption{Comparison of the $\Xstate(v)\to\bstate{}$ dissociation cross sections calculated by \citet{ST98} and \citet{TT01,TT02}
    (ST98 and TT01/02, respectively), with the molecular convergent close-coupling
    (MCCC) calculations~\cite{Scarlett20adndt1,Scarlett20adndt2} for scattering on the $v=0$--4 levels
  of H$_2$.}
  \label{fig:b3Su}
\end{figure}
\begin{figure}[htb]
  \centering
  \includegraphics{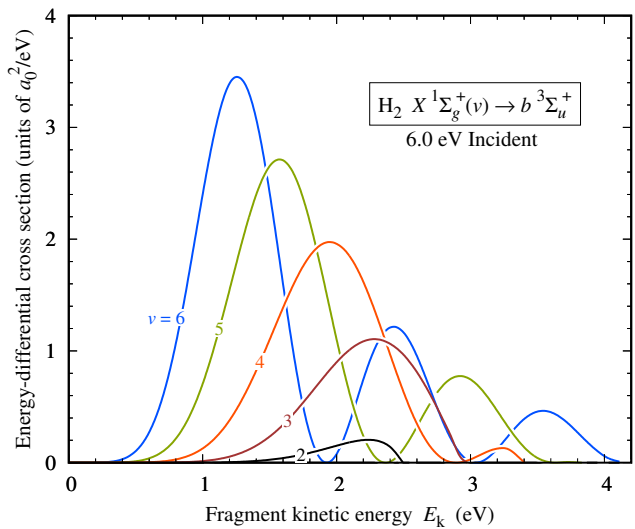}
  \caption{MCCC energy-differential dissociation cross section as a function of the fragment kinetic energy $E_\text{k}$ for
    6.0-eV electrons scattering on the $v=2$--6 vibrational levels of H$_2$.}\label{fig:KER}
\end{figure}

In Fig.~\ref{fig:b3Su} we compare the MCCC and TT01/02 results for dissociation of
H$_2$ in the initial vibrational states $v=0$--4.
We also compare with the earlier $R$-matrix results of \citet{ST98} (ST98), which were performed using the standard AN method.
Both sets of $R$-matrix calculations used the same underlying $T$-matrix elements, calculated previously in an
investigation into H$_2^\minus$ resonances~\cite{Stibbe1998}, so the differences between them is only due to
TT01's alternative cross-section formula. ST98 proposed an ``energy-balancing'' correction to the 
AN method, which slightly modifies the incident energy at which the $T$-matrix elements are evaluated. 
This method was also used by TT01/02 and we have implemented it in the MCCC calculations to ensure consistent 
comparisons between the three sets of calculations. 
The ST98 and MCCC cross sections are in good agreement, with small differences near the cross section maximum
likely due to the use of different electronic scattering models. 
TT01's cross section for the $v=0$ level of H$_2$ is only about 10\% different from ST98's results at the maximum,
and up to around 10~eV the two are essentially the same.
It is perhaps puzzling that the additional factor of $m_\text{H}E_\text{k}/2\pi^4\hbar^2$ in Eq.~(\ref{eq:cs-TT}) has
only a small effect, but we note that the average value of this factor is fortuitously close to 1 
(in Hartree atomic units) in the 0--6~eV range of $E_\text{k}$ corresponding to incident energies up to 10~eV
(see TT01's Fig.~8). 
This coincidence disappears for scattering on excited vibrational levels, where the
formulation of TT01 predicts significantly different results.
The cross sections for
dissociation of excited vibrational levels presented in TT02 are up to a factor of 2 smaller than the MCCC and ST98 
results.
To explain this, in Fig.~\ref{fig:KER} we present the MCCC energy-differential cross section as a 
function of $E_\text{k}$ for 6-eV electrons scattering on the $v=2$--6 levels
of H$_2$, showing that the cross section peaks at progressively smaller values of $E_\text{k}$ as the initial vibrational
level is increased.
In TT01's formalism, the
suppression in this region caused by the smaller value of their additional factor in Eq.~(\ref{eq:cs-TT}) for small $E_\text{k}$
leads to a substantial reduction in the integrated cross section for higher vibrational levels.

\begin{figure}[t!]
  \centering
  \includegraphics{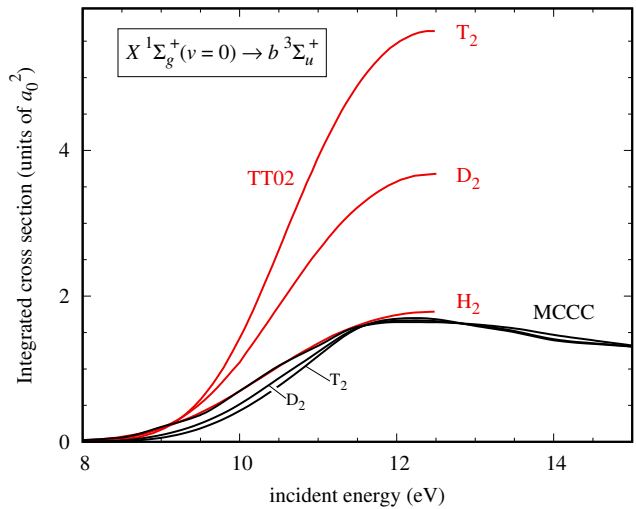}
  \caption{Comparison of the $\Xstate(v=0)\to\bstate{}$ dissociation cross sections calculated by \citet{TT02}
    (TT02), with the molecular convergent close-coupling
  (MCCC) calculations~\cite{Scarlett20adndt1,Scarlett20adndt2} for scattering on H$_2$, D$_2$, and T$_2$.}
  \label{fig:b3Su_iso}
\end{figure}
Further to the different vibrational dependence, the mass factor in Eq.~(\ref{eq:cs-TT}) 
leads to an unusually large 
isotopic dependence in the dissociation cross sections.
In Fig.~\ref{fig:b3Su_iso} we compare the MCCC and TT02 cross sections for dissociation of H$_2$, D$_2$, and 
T$_2$ in the $v=0$ level.
TT02's results for D$_2$ are two times larger than for H$_2$, and for T$_2$ they are three times
larger. 
Our calculations show only a small isotope effect at low energies due to the slightly higher dissociation thresholds of the heavier
targets. Further discussion of the isotope effect in MCCC calculations of low-energy H$_2$ dissociation (including
the mixed isotopologues HD, HT, and DT) can be found in Ref.~\cite{Scarlett17b3}.
TT02 stated that their predicted scaling of cross sections with isotopic mass should be expected
for all dissociative processes.
We suggest that this is purely an artifact of the incorrect formalism they have applied.
Furthermore, there is no substantial isotope effect in the measurements of \citet{AJUD04} for dissociation of 
vibrationally-hot H$_2^+$ and D$_2^+$, which are in good agreement with the previous MCCC calculations~\cite{ZFB14} using the standard AN
method. Additionally, the MCCC calculations are are in good agreement with the measurements of \citet{AJKPV97} for dissociation of
HD$^+$ in its $v=0$ level.

Although we claim that the large isotope effect predicted by TT02 is unphysical, 
higher rates of dissociation for D$_2$ compared to H$_2$ have in fact been observed in
real plasmas~\cite{Heinemann2017,Lamara2006,Mizuochi2012,Rauner2017a,Fantz2013,Wunderlich2016a,Mizuochi2007}.
Each of Refs.~\cite{Heinemann2017,Lamara2006,Mizuochi2012,Rauner2017a,Fantz2013,Wunderlich2016a,Mizuochi2007}
cite TT02's factor-of-two H$_2$/D$_2$ isotope effect as the likely cause
of this, with one~\cite{Mizuochi2007} even repeating the argument that the density of states is 
responsible for the higher rate of dissociation in D$_2$.
Optical-emission spectroscopy measurements have shown a higher ratio of atomic to molecular densities in 
deuterium plasmas than in hydrogen plasmas~\cite{Heinemann2017,Rauner2017a,Fantz2013}, although given the
error bars on the measurements the isotope effect could range from being insubstantial to a factor of 2 difference.
Of course these measurements are indirect and the atomic/molecular densities are governed by a number of 
factors, only one of which is the direct dissociation rate.

\begin{figure}[htb!]
  \centering
  \includegraphics{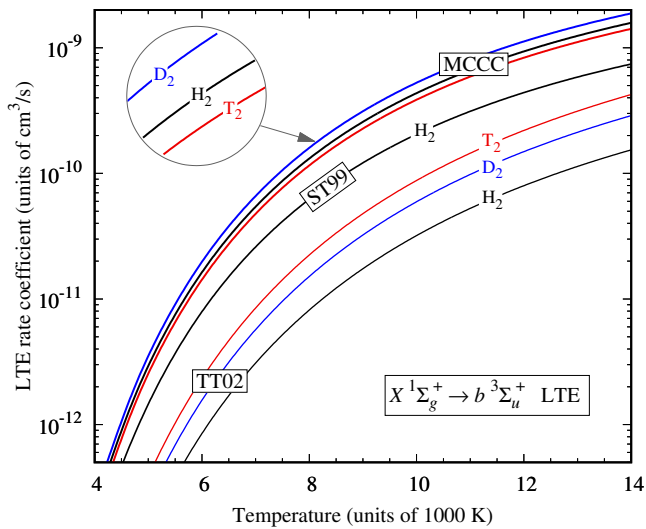}
  \caption{Local thermodynamic equilibrium (LTE) dissociation rate coefficients for H$_2$ (black), D$_2$ (blue), 
    and T$_2$ (red). Comparison is shown between the calculations of \citet{TT02} (TT02), \citet{ST99} (ST99),
    and the present molecular convergent close-coupling (MCCC) calculations.} 
    \label{fig:rates}
\end{figure}

If there is indeed an isotope effect in the rate of dissociation, we argue that it can result from the mass-dependence of the
\Xstate{}-state discrete vibrational spectrum alone, rather than an explicit mass dependence in the cross section. 
Increasing the nuclear mass has two competing effects: 
a shift of the spectrum to lower energies which increases dissociation thresholds, and the appearance of
more bound levels near the dissociative limit of the electronic potential well. 
It is not obvious what isotope dependence (if any) will arise in a plasma from these effects, but for an
approximate idea we can compare the local-thermodynamic-equilibrium (LTE) rate coefficients for different
isotopologues. Since LTE assumes a Maxwellian population of the initial vibrational levels this is a useful way to
investigate the overall isotope effect with excited vibrational levels included.
In Fig.~\ref{fig:rates} we compare the LTE rate coefficients given by \citet{ST99} (ST99) and TT02 with those obtained 
from the MCCC cross sections. The ST99 rates were obtained using the cross sections from ST98.
The MCCC results show an approximately $10\%$ enhancement for dissociation of D$_2$ compared to H$_2$, but
a slightly lower rate for T$_2$. 
The difference between the MCCC rate for H$_2$ and ST99's rate can be explained partly by their somewhat lower
cross sections around the maximum (see Fig.~\ref{fig:b3Su}), and their proper treatment of only the $v=0$--4 levels
(with the remaining cross sections extrapolated).
The much larger difference with the TT02 results is caused by their substantially smaller cross sections for scattering
on excited levels.
Importantly, we have shown that the results of the standard AN formalism are consistent with the experimental
evidence of a higher dissociation rate for D$_2$.

\begin{figure}[t!]
  \centering
  \includegraphics{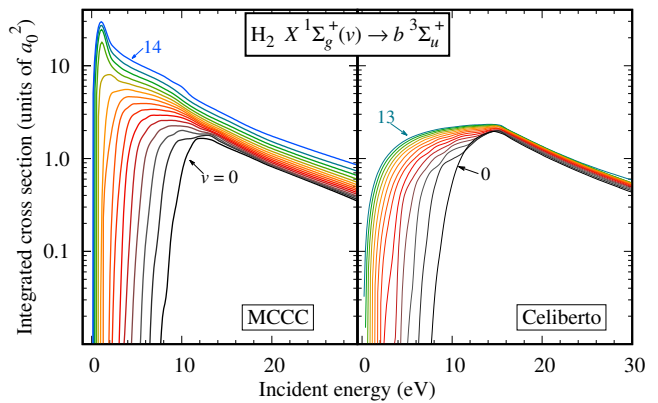}
  \caption{Comparison of the $\Xstate(v)\to\bstate{}$ dissociation cross sections calculated by 
    \citet{CJLCWA01} with the molecular convergent close-coupling (MCCC) 
    calculations~\cite{Scarlett20adndt1}. 
    The cross section increases with initial vibrational level from $v=0$ to 14 (Celiberto et al.\ include only up to $v=13$).
    } 
  \label{fig:b3Su_all_v}
\end{figure}

Aside from TT01/02, the only previous calculations of low-energy dissociation of vibrationally-excited H$_2$ and D$_2$ 
are the semi-classical calculations of \citet{CJLCWA01}, which utilized the Gryzinski method to treat the electronic dynamics
and the standard AN approach with the additional Franck-Condon (FC) approximation to treat the nuclear dynamics. 
The FC approximation assumes the electronic excitation
cross section is independent of $R$, making it is more inaccurate for the higher vibrational levels with more
diffuse wave functions.
In Fig.~\ref{fig:b3Su_all_v} we compare the MCCC \bstate{} cross sections for scattering on all bound vibrational levels
of H$_2$ with the results of \citet{CJLCWA01}. 
It is evident that even with the correct definition of the dissociation cross sections, the approximate methods
utilized by \citet{CJLCWA01} do not correctly model the significant dependence on vibrational level at low 
incident energies. The comparison is essentially the same for D$_2$.

We conclude that the MCCC calculations of low-energy dissociation have treated the nuclear dynamics
correctly (within the AN approximation) and recommend that MCCC cross sections should be used in place of the 
$R$-matrix results~\cite{TT01,TT02,TT02-mixed} in all applications.
The full set of MCCC cross sections for scattering on all bound levels in H$_2$, D$_2$, T$_2$, HD, HT, and DT
are discussed in Refs.~\cite{Scarlett20adndt1,Scarlett20adndt2}, and are available online at \url{mccc-db.org}.
This includes dissociation through the \bstate{} over a much wider range of incident energies than previously 
available, as well as a large number of other inelastic transitions (both bound and dissociative). 
A complete set of rate coefficients for dissociation of each isotopologue will be provided elsewhere, with the 
hope that they will be of interest to those who have previously utilized the data of TT02, and of use
in future models of hydrogenic plasmas.

\begin{acknowledgments}
  This work was supported by the United States Air Force Office of Scientific Research, 
  The Australian Research Council, and resources provided by the Pawsey Supercomputing 
  Centre, with funding from the Australian Government and Government of Western Australia. 
  L.H.S.\ acknowledges the contribution of an Australian Government Research Training Program
  Scholarship, and the support of the Forrest Research Foundation.
  M.C.Z.\ would like to specifically acknowledge LANL’s ASC PEM Atomic Physics Project for its support. 
  LANL is operated by Triad National Security, LLC, for the National Nuclear Security Administration of the 
  U.S. Department of Energy under Contract No. 89233218NCA000001.
\end{acknowledgments}

  \bibliographystyle{apsrev}

\begin{thebibliography}{75}
\expandafter\ifx\csname natexlab\endcsname\relax\def\natexlab#1{#1}\fi
\expandafter\ifx\csname bibnamefont\endcsname\relax
  \def\bibnamefont#1{#1}\fi
\expandafter\ifx\csname bibfnamefont\endcsname\relax
  \def\bibfnamefont#1{#1}\fi
\expandafter\ifx\csname citenamefont\endcsname\relax
  \def\citenamefont#1{#1}\fi
\expandafter\ifx\csname url\endcsname\relax
  \def\url#1{\texttt{#1}}\fi
\expandafter\ifx\csname urlprefix\endcsname\relax\def\urlprefix{URL }\fi
\providecommand{\bibinfo}[2]{#2}
\providecommand{\eprint}[2][]{\url{#2}}

\bibitem[{\citenamefont{Vorobyov et~al.}(2020)\citenamefont{Vorobyov,
  Matsukoba, Omukai, and Guedel}}]{Vorobyov2020}
\bibinfo{author}{\bibfnamefont{E.~I.} \bibnamefont{Vorobyov}},
  \bibinfo{author}{\bibfnamefont{R.}~\bibnamefont{Matsukoba}},
  \bibinfo{author}{\bibfnamefont{K.}~\bibnamefont{Omukai}}, \bibnamefont{and}
  \bibinfo{author}{\bibfnamefont{M.}~\bibnamefont{Guedel}},
  \bibinfo{journal}{Astron. Astrophys.} \textbf{\bibinfo{volume}{638}},
  \bibinfo{pages}{1} (\bibinfo{year}{2020}).

\bibitem[{\citenamefont{Shin et~al.}(2014)\citenamefont{Shin, Park, Jung, Bong,
  Kim, Lee, and Yi}}]{Shin2014}
\bibinfo{author}{\bibfnamefont{C.}~\bibnamefont{Shin}},
  \bibinfo{author}{\bibfnamefont{J.}~\bibnamefont{Park}},
  \bibinfo{author}{\bibfnamefont{J.}~\bibnamefont{Jung}},
  \bibinfo{author}{\bibfnamefont{S.}~\bibnamefont{Bong}},
  \bibinfo{author}{\bibfnamefont{S.}~\bibnamefont{Kim}},
  \bibinfo{author}{\bibfnamefont{Y.~J.} \bibnamefont{Lee}}, \bibnamefont{and}
  \bibinfo{author}{\bibfnamefont{J.}~\bibnamefont{Yi}},
  \bibinfo{journal}{Mater. Res. Bull.} \textbf{\bibinfo{volume}{60}},
  \bibinfo{pages}{895} (\bibinfo{year}{2014}).

\bibitem[{\citenamefont{Broks et~al.}(2005)\citenamefont{Broks, Garloff, and
  {Van Der Mullen}}}]{Broks2005}
\bibinfo{author}{\bibfnamefont{B.}~\bibnamefont{Broks}},
  \bibinfo{author}{\bibfnamefont{K.}~\bibnamefont{Garloff}}, \bibnamefont{and}
  \bibinfo{author}{\bibfnamefont{J.}~\bibnamefont{{Van Der Mullen}}},
  \bibinfo{journal}{Phys. Rev. E - Stat. Nonlinear, Soft Matter Phys.}
  \textbf{\bibinfo{volume}{71}}, \bibinfo{pages}{016401}
  (\bibinfo{year}{2005}).

\bibitem[{\citenamefont{Zammit et~al.}(2017{\natexlab{a}})\citenamefont{Zammit,
  Savage, Fursa, and Bray}}]{ZSFB17}
\bibinfo{author}{\bibfnamefont{M.~C.} \bibnamefont{Zammit}},
  \bibinfo{author}{\bibfnamefont{J.~S.} \bibnamefont{Savage}},
  \bibinfo{author}{\bibfnamefont{D.~V.} \bibnamefont{Fursa}}, \bibnamefont{and}
  \bibinfo{author}{\bibfnamefont{I.}~\bibnamefont{Bray}},
  \bibinfo{journal}{Phys. Rev. A} \textbf{\bibinfo{volume}{95}},
  \bibinfo{pages}{022708} (\bibinfo{year}{2017}{\natexlab{a}}).

\bibitem[{\citenamefont{Zawadzki
  et~al.}(2018{\natexlab{a}})\citenamefont{Zawadzki, Wright, Dolmat, Martin,
  Hargreaves, Fursa, Zammit, Scarlett, Tapley, Savage et~al.}}]{TOF18}
\bibinfo{author}{\bibfnamefont{M.}~\bibnamefont{Zawadzki}},
  \bibinfo{author}{\bibfnamefont{R.}~\bibnamefont{Wright}},
  \bibinfo{author}{\bibfnamefont{G.}~\bibnamefont{Dolmat}},
  \bibinfo{author}{\bibfnamefont{M.~F.} \bibnamefont{Martin}},
  \bibinfo{author}{\bibfnamefont{L.}~\bibnamefont{Hargreaves}},
  \bibinfo{author}{\bibfnamefont{D.~V.} \bibnamefont{Fursa}},
  \bibinfo{author}{\bibfnamefont{M.~C.} \bibnamefont{Zammit}},
  \bibinfo{author}{\bibfnamefont{L.~H.} \bibnamefont{Scarlett}},
  \bibinfo{author}{\bibfnamefont{J.~K.} \bibnamefont{Tapley}},
  \bibinfo{author}{\bibfnamefont{J.~S.} \bibnamefont{Savage}},
  \bibnamefont{et~al.}, \bibinfo{journal}{Phys. Rev. A}
  \textbf{\bibinfo{volume}{97}}, \bibinfo{pages}{050702(R)}
  (\bibinfo{year}{2018}{\natexlab{a}}).

\bibitem[{\citenamefont{Zawadzki
  et~al.}(2018{\natexlab{b}})\citenamefont{Zawadzki, Wright, Dolmat, Martin,
  Diaz, Hargreaves, Coleman, Fursa, Zammit, Scarlett et~al.}}]{TOF18Long}
\bibinfo{author}{\bibfnamefont{M.}~\bibnamefont{Zawadzki}},
  \bibinfo{author}{\bibfnamefont{R.}~\bibnamefont{Wright}},
  \bibinfo{author}{\bibfnamefont{G.}~\bibnamefont{Dolmat}},
  \bibinfo{author}{\bibfnamefont{M.~F.} \bibnamefont{Martin}},
  \bibinfo{author}{\bibfnamefont{B.}~\bibnamefont{Diaz}},
  \bibinfo{author}{\bibfnamefont{L.}~\bibnamefont{Hargreaves}},
  \bibinfo{author}{\bibfnamefont{D.}~\bibnamefont{Coleman}},
  \bibinfo{author}{\bibfnamefont{D.~V.} \bibnamefont{Fursa}},
  \bibinfo{author}{\bibfnamefont{M.~C.} \bibnamefont{Zammit}},
  \bibinfo{author}{\bibfnamefont{L.~H.} \bibnamefont{Scarlett}},
  \bibnamefont{et~al.}, \bibinfo{journal}{Phys. Rev. A}
  \textbf{\bibinfo{volume}{98}}, \bibinfo{pages}{062704}
  (\bibinfo{year}{2018}{\natexlab{b}}).

\bibitem[{\citenamefont{Meltzer et~al.}(2020)\citenamefont{Meltzer, Tennyson,
  Masin, Zammit, Scarlett, Fursa, and Bray}}]{Meltzer2020}
\bibinfo{author}{\bibfnamefont{T.}~\bibnamefont{Meltzer}},
  \bibinfo{author}{\bibfnamefont{J.}~\bibnamefont{Tennyson}},
  \bibinfo{author}{\bibfnamefont{Z.}~\bibnamefont{Masin}},
  \bibinfo{author}{\bibfnamefont{M.~C.} \bibnamefont{Zammit}},
  \bibinfo{author}{\bibfnamefont{L.~H.} \bibnamefont{Scarlett}},
  \bibinfo{author}{\bibfnamefont{D.~V.} \bibnamefont{Fursa}}, \bibnamefont{and}
  \bibinfo{author}{\bibfnamefont{I.}~\bibnamefont{Bray}}, \bibinfo{journal}{J.
  Phys. B} \textbf{\bibinfo{volume}{53}}, \bibinfo{pages}{14}
  (\bibinfo{year}{2020}).

\bibitem[{\citenamefont{Fliflet and McKoy}(1980)}]{FM80}
\bibinfo{author}{\bibfnamefont{A.~W.} \bibnamefont{Fliflet}} \bibnamefont{and}
  \bibinfo{author}{\bibfnamefont{V.}~\bibnamefont{McKoy}},
  \bibinfo{journal}{Phys. Rev. A} \textbf{\bibinfo{volume}{21}},
  \bibinfo{pages}{1863} (\bibinfo{year}{1980}).

\bibitem[{\citenamefont{Schneider and Collins}(1985)}]{SC85}
\bibinfo{author}{\bibfnamefont{B.~I.} \bibnamefont{Schneider}}
  \bibnamefont{and} \bibinfo{author}{\bibfnamefont{L.~A.}
  \bibnamefont{Collins}}, \bibinfo{journal}{J. Phys. B At. Mol. Opt. Phys.}
  \textbf{\bibinfo{volume}{18}}, \bibinfo{pages}{L857} (\bibinfo{year}{1985}),
  ISSN \bibinfo{issn}{09534075}.

\bibitem[{\citenamefont{Baluja et~al.}(1985)\citenamefont{Baluja, Noble, and
  Tennyson}}]{Baluja1985a}
\bibinfo{author}{\bibfnamefont{K.~L.} \bibnamefont{Baluja}},
  \bibinfo{author}{\bibfnamefont{C.~J.} \bibnamefont{Noble}}, \bibnamefont{and}
  \bibinfo{author}{\bibfnamefont{J.}~\bibnamefont{Tennyson}},
  \bibinfo{journal}{J. Phys. B At. Mol. Phys.} \textbf{\bibinfo{volume}{18}}
  (\bibinfo{year}{1985}).

\bibitem[{\citenamefont{Lima et~al.}(1984)\citenamefont{Lima, Gibson,
  Takatsuka, and McKoy}}]{Lima1984}
\bibinfo{author}{\bibfnamefont{M.~A.~P.} \bibnamefont{Lima}},
  \bibinfo{author}{\bibfnamefont{T.~L.} \bibnamefont{Gibson}},
  \bibinfo{author}{\bibfnamefont{K.}~\bibnamefont{Takatsuka}},
  \bibnamefont{and} \bibinfo{author}{\bibfnamefont{V.}~\bibnamefont{McKoy}},
  \bibinfo{journal}{Phys. Rev. A} \textbf{\bibinfo{volume}{30}},
  \bibinfo{pages}{1741} (\bibinfo{year}{1984}), ISSN \bibinfo{issn}{10502947}.

\bibitem[{\citenamefont{Lima et~al.}(1985)\citenamefont{Lima, Gibson, Huo, and
  McKoy}}]{Lima1985}
\bibinfo{author}{\bibfnamefont{M.~A.} \bibnamefont{Lima}},
  \bibinfo{author}{\bibfnamefont{T.~L.} \bibnamefont{Gibson}},
  \bibinfo{author}{\bibfnamefont{W.~M.} \bibnamefont{Huo}}, \bibnamefont{and}
  \bibinfo{author}{\bibfnamefont{V.}~\bibnamefont{McKoy}}, \bibinfo{journal}{J.
  Phys. B At. Mol. Phys.} \textbf{\bibinfo{volume}{18}} (\bibinfo{year}{1985}).

\bibitem[{\citenamefont{Rescigno and Schneider}(1988)}]{RS88}
\bibinfo{author}{\bibfnamefont{T.~N.} \bibnamefont{Rescigno}} \bibnamefont{and}
  \bibinfo{author}{\bibfnamefont{B.~I.} \bibnamefont{Schneider}},
  \bibinfo{journal}{J. Phys. B At. Mol. Phys.} \textbf{\bibinfo{volume}{21}},
  \bibinfo{pages}{L691} (\bibinfo{year}{1988}).

\bibitem[{\citenamefont{Branchett et~al.}(1990)\citenamefont{Branchett,
  Tennyson, and Morgan}}]{BTM90}
\bibinfo{author}{\bibfnamefont{S.~E.} \bibnamefont{Branchett}},
  \bibinfo{author}{\bibfnamefont{J.}~\bibnamefont{Tennyson}}, \bibnamefont{and}
  \bibinfo{author}{\bibfnamefont{L.~A.} \bibnamefont{Morgan}},
  \bibinfo{journal}{J. Phys. B At. Mol. Opt. Phys.}
  \textbf{\bibinfo{volume}{23}}, \bibinfo{pages}{4625} (\bibinfo{year}{1990}).

\bibitem[{\citenamefont{Stibbe and Tennyson}(1998{\natexlab{a}})}]{ST98}
\bibinfo{author}{\bibfnamefont{D.~T.} \bibnamefont{Stibbe}} \bibnamefont{and}
  \bibinfo{author}{\bibfnamefont{J.}~\bibnamefont{Tennyson}},
  \bibinfo{journal}{New J. Phys.} \textbf{\bibinfo{volume}{1}},
  \bibinfo{pages}{2} (\bibinfo{year}{1998}{\natexlab{a}}).

\bibitem[{\citenamefont{Celiberto et~al.}(2001)\citenamefont{Celiberto, Jenev,
  Laricchiuta, Capitelli, Wadehra, and Atems}}]{CJLCWA01}
\bibinfo{author}{\bibfnamefont{R.}~\bibnamefont{Celiberto}},
  \bibinfo{author}{\bibfnamefont{R.}~\bibnamefont{Jenev}},
  \bibinfo{author}{\bibfnamefont{A.}~\bibnamefont{Laricchiuta}},
  \bibinfo{author}{\bibfnamefont{M.}~\bibnamefont{Capitelli}},
  \bibinfo{author}{\bibfnamefont{J.}~\bibnamefont{Wadehra}}, \bibnamefont{and}
  \bibinfo{author}{\bibfnamefont{D.}~\bibnamefont{Atems}},
  \bibinfo{journal}{At. Data Nucl. Data Tables} \textbf{\bibinfo{volume}{77}},
  \bibinfo{pages}{161} (\bibinfo{year}{2001}).

\bibitem[{\citenamefont{Trevisan and Tennyson}(2001)}]{TT01}
\bibinfo{author}{\bibfnamefont{C.~S.} \bibnamefont{Trevisan}} \bibnamefont{and}
  \bibinfo{author}{\bibfnamefont{J.}~\bibnamefont{Tennyson}},
  \bibinfo{journal}{J. Phys. B At. Mol. Opt. Phys.}
  \textbf{\bibinfo{volume}{34}}, \bibinfo{pages}{2935} (\bibinfo{year}{2001}).

\bibitem[{\citenamefont{da~Costa et~al.}(2005)\citenamefont{da~Costa,
  da~Paix{\~{a}}o, and Lima}}]{CPL05}
\bibinfo{author}{\bibfnamefont{R.~F.} \bibnamefont{da~Costa}},
  \bibinfo{author}{\bibfnamefont{F.~J.} \bibnamefont{da~Paix{\~{a}}o}},
  \bibnamefont{and} \bibinfo{author}{\bibfnamefont{M.~A.~P.}
  \bibnamefont{Lima}}, \bibinfo{journal}{J. Phys. B At. Mol. Opt. Phys.}
  \textbf{\bibinfo{volume}{38}}, \bibinfo{pages}{4363} (\bibinfo{year}{2005}).

\bibitem[{\citenamefont{Gorfinkiel and Tennyson}(2005)}]{GT05}
\bibinfo{author}{\bibfnamefont{J.~D.} \bibnamefont{Gorfinkiel}}
  \bibnamefont{and} \bibinfo{author}{\bibfnamefont{J.}~\bibnamefont{Tennyson}},
  \bibinfo{journal}{J. Phys. B} \textbf{\bibinfo{volume}{38}},
  \bibinfo{pages}{1607} (\bibinfo{year}{2005}).

\bibitem[{\citenamefont{Zammit et~al.}(2014)\citenamefont{Zammit, Fursa, and
  Bray}}]{ZFB14}
\bibinfo{author}{\bibfnamefont{M.~C.} \bibnamefont{Zammit}},
  \bibinfo{author}{\bibfnamefont{D.~V.} \bibnamefont{Fursa}}, \bibnamefont{and}
  \bibinfo{author}{\bibfnamefont{I.}~\bibnamefont{Bray}},
  \bibinfo{journal}{Phys. Rev. A} \textbf{\bibinfo{volume}{90}},
  \bibinfo{pages}{022711} (\bibinfo{year}{2014}).

\bibitem[{\citenamefont{Scarlett
  et~al.}(2017{\natexlab{a}})\citenamefont{Scarlett, Zammit, Fursa, and
  Bray}}]{Scarlett17}
\bibinfo{author}{\bibfnamefont{L.~H.} \bibnamefont{Scarlett}},
  \bibinfo{author}{\bibfnamefont{M.~C.} \bibnamefont{Zammit}},
  \bibinfo{author}{\bibfnamefont{D.~V.} \bibnamefont{Fursa}}, \bibnamefont{and}
  \bibinfo{author}{\bibfnamefont{I.}~\bibnamefont{Bray}},
  \bibinfo{journal}{Phys. Rev. A} \textbf{\bibinfo{volume}{96}},
  \bibinfo{pages}{022706} (\bibinfo{year}{2017}{\natexlab{a}}).

\bibitem[{\citenamefont{Scarlett
  et~al.}(2017{\natexlab{b}})\citenamefont{Scarlett, Tapley, Fursa, Zammit,
  Savage, and Bray}}]{Scarlett17b3}
\bibinfo{author}{\bibfnamefont{L.~H.} \bibnamefont{Scarlett}},
  \bibinfo{author}{\bibfnamefont{J.~K.} \bibnamefont{Tapley}},
  \bibinfo{author}{\bibfnamefont{D.~V.} \bibnamefont{Fursa}},
  \bibinfo{author}{\bibfnamefont{M.~C.} \bibnamefont{Zammit}},
  \bibinfo{author}{\bibfnamefont{J.~S.} \bibnamefont{Savage}},
  \bibnamefont{and} \bibinfo{author}{\bibfnamefont{I.}~\bibnamefont{Bray}},
  \bibinfo{journal}{Phys. Rev. A} \textbf{\bibinfo{volume}{96}},
  \bibinfo{pages}{062708} (\bibinfo{year}{2017}{\natexlab{b}}).

\bibitem[{\citenamefont{Tapley et~al.}(2018)\citenamefont{Tapley, Scarlett,
  Savage, Fursa, Zammit, and Bray}}]{Tapley18DE}
\bibinfo{author}{\bibfnamefont{J.~K.} \bibnamefont{Tapley}},
  \bibinfo{author}{\bibfnamefont{L.~H.} \bibnamefont{Scarlett}},
  \bibinfo{author}{\bibfnamefont{J.~S.} \bibnamefont{Savage}},
  \bibinfo{author}{\bibfnamefont{D.~V.} \bibnamefont{Fursa}},
  \bibinfo{author}{\bibfnamefont{M.~C.} \bibnamefont{Zammit}},
  \bibnamefont{and} \bibinfo{author}{\bibfnamefont{I.}~\bibnamefont{Bray}},
  \bibinfo{journal}{Phys. Rev. A} \textbf{\bibinfo{volume}{98}},
  \bibinfo{pages}{032701} (\bibinfo{year}{2018}).

\bibitem[{\citenamefont{Laricchiuta et~al.}(2004)\citenamefont{Laricchiuta,
  Celiberto, and Janev}}]{LaricchiutaCJ04}
\bibinfo{author}{\bibfnamefont{A.}~\bibnamefont{Laricchiuta}},
  \bibinfo{author}{\bibfnamefont{R.}~\bibnamefont{Celiberto}},
  \bibnamefont{and} \bibinfo{author}{\bibfnamefont{R.~K.} \bibnamefont{Janev}},
  \bibinfo{journal}{Phys. Rev. A} \textbf{\bibinfo{volume}{69}},
  \bibinfo{pages}{22706} (\bibinfo{year}{2004}).

\bibitem[{\citenamefont{Lane}(1980)}]{Lane80}
\bibinfo{author}{\bibfnamefont{N.~F.} \bibnamefont{Lane}},
  \bibinfo{journal}{Rev. Mod. Phys.} \textbf{\bibinfo{volume}{52}},
  \bibinfo{pages}{29} (\bibinfo{year}{1980}).

\bibitem[{\citenamefont{Yoon et~al.}(2010)\citenamefont{Yoon, Kim, Kwon, Song,
  Chang, Kim, Kumar, and Lee}}]{Yoon2010}
\bibinfo{author}{\bibfnamefont{J.~S.} \bibnamefont{Yoon}},
  \bibinfo{author}{\bibfnamefont{Y.~W.} \bibnamefont{Kim}},
  \bibinfo{author}{\bibfnamefont{D.~C.} \bibnamefont{Kwon}},
  \bibinfo{author}{\bibfnamefont{M.~Y.} \bibnamefont{Song}},
  \bibinfo{author}{\bibfnamefont{W.~S.} \bibnamefont{Chang}},
  \bibinfo{author}{\bibfnamefont{C.~G.} \bibnamefont{Kim}},
  \bibinfo{author}{\bibfnamefont{V.}~\bibnamefont{Kumar}}, \bibnamefont{and}
  \bibinfo{author}{\bibfnamefont{B.~J.} \bibnamefont{Lee}},
  \bibinfo{journal}{Reports Prog. Phys.} \textbf{\bibinfo{volume}{73}}
  (\bibinfo{year}{2010}).

\bibitem[{\citenamefont{Yoon et~al.}(2014)\citenamefont{Yoon, Song, Kwon, Choi,
  Kim, and Kumar}}]{Yoon2014}
\bibinfo{author}{\bibfnamefont{J.-S.} \bibnamefont{Yoon}},
  \bibinfo{author}{\bibfnamefont{M.-Y.} \bibnamefont{Song}},
  \bibinfo{author}{\bibfnamefont{D.-C.} \bibnamefont{Kwon}},
  \bibinfo{author}{\bibfnamefont{H.}~\bibnamefont{Choi}},
  \bibinfo{author}{\bibfnamefont{C.-G.} \bibnamefont{Kim}}, \bibnamefont{and}
  \bibinfo{author}{\bibfnamefont{V.}~\bibnamefont{Kumar}},
  \bibinfo{journal}{Phys. Rep.} \textbf{\bibinfo{volume}{543}},
  \bibinfo{pages}{199} (\bibinfo{year}{2014}).

\bibitem[{\citenamefont{Trevisan and Tennyson}(2002{\natexlab{a}})}]{TT02}
\bibinfo{author}{\bibfnamefont{C.}~\bibnamefont{Trevisan}} \bibnamefont{and}
  \bibinfo{author}{\bibfnamefont{J.}~\bibnamefont{Tennyson}},
  \bibinfo{journal}{Plasma Phys. Contr. F.} \textbf{\bibinfo{volume}{44}},
  \bibinfo{pages}{1263} (\bibinfo{year}{2002}{\natexlab{a}}).

\bibitem[{\citenamefont{Trevisan and
  Tennyson}(2002{\natexlab{b}})}]{TT02-mixed}
\bibinfo{author}{\bibfnamefont{C.~S.} \bibnamefont{Trevisan}} \bibnamefont{and}
  \bibinfo{author}{\bibfnamefont{J.}~\bibnamefont{Tennyson}},
  \bibinfo{journal}{Plasma Phys. Control. Fusion}
  \textbf{\bibinfo{volume}{44}}, \bibinfo{pages}{2217}
  (\bibinfo{year}{2002}{\natexlab{b}}).

\bibitem[{\citenamefont{Scarlett
  et~al.}(2020{\natexlab{a}})\citenamefont{Scarlett, Fursa, Zammit, Bray,
  Ralchenko, and Davie}}]{Scarlett20adndt1}
\bibinfo{author}{\bibfnamefont{L.~H.} \bibnamefont{Scarlett}},
  \bibinfo{author}{\bibfnamefont{D.~V.} \bibnamefont{Fursa}},
  \bibinfo{author}{\bibfnamefont{M.~C.} \bibnamefont{Zammit}},
  \bibinfo{author}{\bibfnamefont{I.}~\bibnamefont{Bray}},
  \bibinfo{author}{\bibfnamefont{Y.}~\bibnamefont{Ralchenko}},
  \bibnamefont{and} \bibinfo{author}{\bibfnamefont{K.~D.} \bibnamefont{Davie}},
  \bibinfo{journal}{At. Data Nucl. Data Tables} p. \bibinfo{pages}{101361}
  (\bibinfo{year}{2020}{\natexlab{a}}).

\bibitem[{\citenamefont{Scarlett
  et~al.}(2020{\natexlab{b}})\citenamefont{Scarlett, Fursa, Zammit, Bray, and
  Ralchenko}}]{Scarlett20adndt2}
\bibinfo{author}{\bibfnamefont{L.~H.} \bibnamefont{Scarlett}},
  \bibinfo{author}{\bibfnamefont{D.~V.} \bibnamefont{Fursa}},
  \bibinfo{author}{\bibfnamefont{M.~C.} \bibnamefont{Zammit}},
  \bibinfo{author}{\bibfnamefont{I.}~\bibnamefont{Bray}}, \bibnamefont{and}
  \bibinfo{author}{\bibfnamefont{Y.}~\bibnamefont{Ralchenko}},
  \bibinfo{journal}{At. Data Nucl. Data Tables}
  \textbf{\bibinfo{volume}{submitted}} (\bibinfo{year}{2020}{\natexlab{b}}).

\bibitem[{\citenamefont{Yoon et~al.}(2008)\citenamefont{Yoon, Song, Han, Hwang,
  Chang, Lee, and Itikawa}}]{Yoon2008}
\bibinfo{author}{\bibfnamefont{J.~S.} \bibnamefont{Yoon}},
  \bibinfo{author}{\bibfnamefont{M.~Y.} \bibnamefont{Song}},
  \bibinfo{author}{\bibfnamefont{J.~M.} \bibnamefont{Han}},
  \bibinfo{author}{\bibfnamefont{S.~H.} \bibnamefont{Hwang}},
  \bibinfo{author}{\bibfnamefont{W.~S.} \bibnamefont{Chang}},
  \bibinfo{author}{\bibfnamefont{B.}~\bibnamefont{Lee}}, \bibnamefont{and}
  \bibinfo{author}{\bibfnamefont{Y.}~\bibnamefont{Itikawa}},
  \bibinfo{journal}{J. Phys. Chem. Ref. Data} \textbf{\bibinfo{volume}{37}},
  \bibinfo{pages}{913} (\bibinfo{year}{2008}).

\bibitem[{\citenamefont{Raju}(2006)}]{Raju2006}
\bibinfo{author}{\bibfnamefont{G.~G.} \bibnamefont{Raju}}, in
  \emph{\bibinfo{booktitle}{Gaseous Electron. Theory Pract.}}
  (\bibinfo{publisher}{CRC Press}, \bibinfo{address}{Boca Raton, FL},
  \bibinfo{year}{2006}), chap.~\bibinfo{chapter}{4}.

\bibitem[{\citenamefont{Ziegler}(2018)}]{Ziegler2018}
\bibinfo{author}{\bibfnamefont{U.}~\bibnamefont{Ziegler}},
  \bibinfo{journal}{Astron. Astrophys.} \textbf{\bibinfo{volume}{620}}
  (\bibinfo{year}{2018}).

\bibitem[{\citenamefont{Sormani et~al.}(2018)\citenamefont{Sormani, Tre{\ss},
  Ridley, Glover, Klessen, Binney, Magorrian, and Smith}}]{Sormani2018}
\bibinfo{author}{\bibfnamefont{M.~C.} \bibnamefont{Sormani}},
  \bibinfo{author}{\bibfnamefont{R.~G.} \bibnamefont{Tre{\ss}}},
  \bibinfo{author}{\bibfnamefont{M.}~\bibnamefont{Ridley}},
  \bibinfo{author}{\bibfnamefont{S.~C.} \bibnamefont{Glover}},
  \bibinfo{author}{\bibfnamefont{R.~S.} \bibnamefont{Klessen}},
  \bibinfo{author}{\bibfnamefont{J.}~\bibnamefont{Binney}},
  \bibinfo{author}{\bibfnamefont{J.}~\bibnamefont{Magorrian}},
  \bibnamefont{and} \bibinfo{author}{\bibfnamefont{R.}~\bibnamefont{Smith}},
  \bibinfo{journal}{Mon. Not. R. Astron. Soc.} \textbf{\bibinfo{volume}{475}},
  \bibinfo{pages}{2383} (\bibinfo{year}{2018}).

\bibitem[{\citenamefont{Smith et~al.}(2017)\citenamefont{Smith, Bryan, Glover,
  Goldbaum, Turk, Regan, Wise, Schive, Abel, Emerick et~al.}}]{Smith2017}
\bibinfo{author}{\bibfnamefont{B.~D.} \bibnamefont{Smith}},
  \bibinfo{author}{\bibfnamefont{G.~L.} \bibnamefont{Bryan}},
  \bibinfo{author}{\bibfnamefont{S.~C.} \bibnamefont{Glover}},
  \bibinfo{author}{\bibfnamefont{N.~J.} \bibnamefont{Goldbaum}},
  \bibinfo{author}{\bibfnamefont{M.~J.} \bibnamefont{Turk}},
  \bibinfo{author}{\bibfnamefont{J.}~\bibnamefont{Regan}},
  \bibinfo{author}{\bibfnamefont{J.~H.} \bibnamefont{Wise}},
  \bibinfo{author}{\bibfnamefont{H.~Y.} \bibnamefont{Schive}},
  \bibinfo{author}{\bibfnamefont{T.}~\bibnamefont{Abel}},
  \bibinfo{author}{\bibfnamefont{A.}~\bibnamefont{Emerick}},
  \bibnamefont{et~al.}, \bibinfo{journal}{Mon. Not. R. Astron. Soc.}
  \textbf{\bibinfo{volume}{466}}, \bibinfo{pages}{2217} (\bibinfo{year}{2017}).

\bibitem[{\citenamefont{Schleicher et~al.}(2008)\citenamefont{Schleicher,
  Galli, Palla, Camenzind, Klessen, Bartelmann, and Glover}}]{Schleicher2008}
\bibinfo{author}{\bibfnamefont{D.~R.} \bibnamefont{Schleicher}},
  \bibinfo{author}{\bibfnamefont{D.}~\bibnamefont{Galli}},
  \bibinfo{author}{\bibfnamefont{F.}~\bibnamefont{Palla}},
  \bibinfo{author}{\bibfnamefont{M.}~\bibnamefont{Camenzind}},
  \bibinfo{author}{\bibfnamefont{R.~S.} \bibnamefont{Klessen}},
  \bibinfo{author}{\bibfnamefont{M.}~\bibnamefont{Bartelmann}},
  \bibnamefont{and} \bibinfo{author}{\bibfnamefont{S.~C.}
  \bibnamefont{Glover}}, \bibinfo{journal}{Astron. Astrophys.}
  \textbf{\bibinfo{volume}{490}}, \bibinfo{pages}{521} (\bibinfo{year}{2008}).

\bibitem[{\citenamefont{Richings et~al.}(2014)\citenamefont{Richings, Schaye,
  and Oppenheimer}}]{Richings2014}
\bibinfo{author}{\bibfnamefont{A.~J.} \bibnamefont{Richings}},
  \bibinfo{author}{\bibfnamefont{J.}~\bibnamefont{Schaye}}, \bibnamefont{and}
  \bibinfo{author}{\bibfnamefont{B.~D.} \bibnamefont{Oppenheimer}},
  \bibinfo{journal}{Mon. Not. R. Astron. Soc.} \textbf{\bibinfo{volume}{440}},
  \bibinfo{pages}{3349} (\bibinfo{year}{2014}).

\bibitem[{\citenamefont{Micic et~al.}(2012)\citenamefont{Micic, Glover,
  Federrath, and Klessen}}]{Micic2012}
\bibinfo{author}{\bibfnamefont{M.}~\bibnamefont{Micic}},
  \bibinfo{author}{\bibfnamefont{S.~C.} \bibnamefont{Glover}},
  \bibinfo{author}{\bibfnamefont{C.}~\bibnamefont{Federrath}},
  \bibnamefont{and} \bibinfo{author}{\bibfnamefont{R.~S.}
  \bibnamefont{Klessen}}, \bibinfo{journal}{Mon. Not. R. Astron. Soc.}
  \textbf{\bibinfo{volume}{421}}, \bibinfo{pages}{2531} (\bibinfo{year}{2012}).

\bibitem[{\citenamefont{Micic et~al.}(2013)\citenamefont{Micic, Glover,
  Banerjee, and Klessen}}]{Micic2013}
\bibinfo{author}{\bibfnamefont{M.}~\bibnamefont{Micic}},
  \bibinfo{author}{\bibfnamefont{S.~C.} \bibnamefont{Glover}},
  \bibinfo{author}{\bibfnamefont{R.}~\bibnamefont{Banerjee}}, \bibnamefont{and}
  \bibinfo{author}{\bibfnamefont{R.~S.} \bibnamefont{Klessen}},
  \bibinfo{journal}{Mon. Not. R. Astron. Soc.} \textbf{\bibinfo{volume}{432}},
  \bibinfo{pages}{626} (\bibinfo{year}{2013}).

\bibitem[{\citenamefont{Matsukoba et~al.}(2019)\citenamefont{Matsukoba,
  Takahashi, Sugimura, and Omukai}}]{Matsukoba2019}
\bibinfo{author}{\bibfnamefont{R.}~\bibnamefont{Matsukoba}},
  \bibinfo{author}{\bibfnamefont{S.~Z.} \bibnamefont{Takahashi}},
  \bibinfo{author}{\bibfnamefont{K.}~\bibnamefont{Sugimura}}, \bibnamefont{and}
  \bibinfo{author}{\bibfnamefont{K.}~\bibnamefont{Omukai}},
  \bibinfo{journal}{Mon. Not. R. Astron. Soc.} \textbf{\bibinfo{volume}{484}},
  \bibinfo{pages}{2605} (\bibinfo{year}{2019}).

\bibitem[{\citenamefont{MacKey et~al.}(2019)\citenamefont{MacKey, Walch,
  Seifried, Glover, W{\"{u}}nsch, and Aharonian}}]{MacKey2019}
\bibinfo{author}{\bibfnamefont{J.}~\bibnamefont{MacKey}},
  \bibinfo{author}{\bibfnamefont{S.}~\bibnamefont{Walch}},
  \bibinfo{author}{\bibfnamefont{D.}~\bibnamefont{Seifried}},
  \bibinfo{author}{\bibfnamefont{S.~C.} \bibnamefont{Glover}},
  \bibinfo{author}{\bibfnamefont{R.}~\bibnamefont{W{\"{u}}nsch}},
  \bibnamefont{and}
  \bibinfo{author}{\bibfnamefont{F.}~\bibnamefont{Aharonian}},
  \bibinfo{journal}{Mon. Not. R. Astron. Soc.} \textbf{\bibinfo{volume}{486}},
  \bibinfo{pages}{1094} (\bibinfo{year}{2019}).

\bibitem[{\citenamefont{Glover and Savin}(2009)}]{Glover2009}
\bibinfo{author}{\bibfnamefont{S.~C.} \bibnamefont{Glover}} \bibnamefont{and}
  \bibinfo{author}{\bibfnamefont{D.~W.} \bibnamefont{Savin}},
  \bibinfo{journal}{Mon. Not. R. Astron. Soc.} \textbf{\bibinfo{volume}{393}},
  \bibinfo{pages}{911} (\bibinfo{year}{2009}).

\bibitem[{\citenamefont{Glover and Abel}(2008)}]{Glover2008}
\bibinfo{author}{\bibfnamefont{S.~C.} \bibnamefont{Glover}} \bibnamefont{and}
  \bibinfo{author}{\bibfnamefont{T.}~\bibnamefont{Abel}},
  \bibinfo{journal}{Mon. Not. R. Astron. Soc.} \textbf{\bibinfo{volume}{388}},
  \bibinfo{pages}{1627} (\bibinfo{year}{2008}).

\bibitem[{\citenamefont{Coppola et~al.}(2013)\citenamefont{Coppola, Galli,
  Palla, Longo, and Chluba}}]{Coppola2013}
\bibinfo{author}{\bibfnamefont{C.~M.} \bibnamefont{Coppola}},
  \bibinfo{author}{\bibfnamefont{D.}~\bibnamefont{Galli}},
  \bibinfo{author}{\bibfnamefont{F.}~\bibnamefont{Palla}},
  \bibinfo{author}{\bibfnamefont{S.}~\bibnamefont{Longo}}, \bibnamefont{and}
  \bibinfo{author}{\bibfnamefont{J.}~\bibnamefont{Chluba}},
  \bibinfo{journal}{Mon. Not. R. Astron. Soc.} \textbf{\bibinfo{volume}{434}},
  \bibinfo{pages}{114} (\bibinfo{year}{2013}).

\bibitem[{\citenamefont{Clark et~al.}(2019)\citenamefont{Clark, Glover, Ragan,
  and Duarte-Cabral}}]{Clark2019}
\bibinfo{author}{\bibfnamefont{P.~C.} \bibnamefont{Clark}},
  \bibinfo{author}{\bibfnamefont{S.~C.} \bibnamefont{Glover}},
  \bibinfo{author}{\bibfnamefont{S.~E.} \bibnamefont{Ragan}}, \bibnamefont{and}
  \bibinfo{author}{\bibfnamefont{A.}~\bibnamefont{Duarte-Cabral}},
  \bibinfo{journal}{Mon. Not. R. Astron. Soc.} \textbf{\bibinfo{volume}{486}},
  \bibinfo{pages}{4622} (\bibinfo{year}{2019}).

\bibitem[{\citenamefont{Clark et~al.}(2011)\citenamefont{Clark, Glover,
  Klessen, and Bromm}}]{Clark2011}
\bibinfo{author}{\bibfnamefont{P.~C.} \bibnamefont{Clark}},
  \bibinfo{author}{\bibfnamefont{S.~C.} \bibnamefont{Glover}},
  \bibinfo{author}{\bibfnamefont{R.~S.} \bibnamefont{Klessen}},
  \bibnamefont{and} \bibinfo{author}{\bibfnamefont{V.}~\bibnamefont{Bromm}},
  \bibinfo{journal}{Astrophys. J.} \textbf{\bibinfo{volume}{727}},
  \bibinfo{pages}{1} (\bibinfo{year}{2011}).

\bibitem[{\citenamefont{Moravej et~al.}(2004)\citenamefont{Moravej, Babayan,
  Nowling, Yang, and Hicks}}]{Moravej2004}
\bibinfo{author}{\bibfnamefont{M.}~\bibnamefont{Moravej}},
  \bibinfo{author}{\bibfnamefont{S.~E.} \bibnamefont{Babayan}},
  \bibinfo{author}{\bibfnamefont{G.~R.} \bibnamefont{Nowling}},
  \bibinfo{author}{\bibfnamefont{X.}~\bibnamefont{Yang}}, \bibnamefont{and}
  \bibinfo{author}{\bibfnamefont{R.~F.} \bibnamefont{Hicks}},
  \bibinfo{journal}{Plasma Sources Sci. Technol.}
  \textbf{\bibinfo{volume}{13}}, \bibinfo{pages}{8} (\bibinfo{year}{2004}).

\bibitem[{\citenamefont{Scarlett et~al.}(2019)\citenamefont{Scarlett, Savage,
  Fursa, Zammit, and Bray}}]{Scarlett19diss}
\bibinfo{author}{\bibfnamefont{L.~H.} \bibnamefont{Scarlett}},
  \bibinfo{author}{\bibfnamefont{J.~S.} \bibnamefont{Savage}},
  \bibinfo{author}{\bibfnamefont{D.~V.} \bibnamefont{Fursa}},
  \bibinfo{author}{\bibfnamefont{M.~C.} \bibnamefont{Zammit}},
  \bibnamefont{and} \bibinfo{author}{\bibfnamefont{I.}~\bibnamefont{Bray}},
  \bibinfo{journal}{Atoms} \textbf{\bibinfo{volume}{7}}, \bibinfo{pages}{75}
  (\bibinfo{year}{2019}).

\bibitem[{\citenamefont{Cohen-Tannoudji
  et~al.}(1978)\citenamefont{Cohen-Tannoudji, Diu, and Laloe}}]{ClaudeCohenQM}
\bibinfo{author}{\bibfnamefont{C.}~\bibnamefont{Cohen-Tannoudji}},
  \bibinfo{author}{\bibfnamefont{B.}~\bibnamefont{Diu}}, \bibnamefont{and}
  \bibinfo{author}{\bibfnamefont{F.}~\bibnamefont{Laloe}},
  \emph{\bibinfo{title}{{Quantum Mechanics. Volume II}}}
  (\bibinfo{publisher}{John Wiley and Sons}, \bibinfo{year}{1978}).

\bibitem[{\citenamefont{Celiberto et~al.}(1989)\citenamefont{Celiberto,
  Cacciatore, Capitelli, and Gorse}}]{Celiberto1989}
\bibinfo{author}{\bibfnamefont{R.}~\bibnamefont{Celiberto}},
  \bibinfo{author}{\bibfnamefont{M.}~\bibnamefont{Cacciatore}},
  \bibinfo{author}{\bibfnamefont{M.}~\bibnamefont{Capitelli}},
  \bibnamefont{and} \bibinfo{author}{\bibfnamefont{C.}~\bibnamefont{Gorse}},
  \bibinfo{journal}{Chem. Phys.} \textbf{\bibinfo{volume}{133}},
  \bibinfo{pages}{355} (\bibinfo{year}{1989}).

\bibitem[{\citenamefont{Celiberto et~al.}(1994)\citenamefont{Celiberto,
  Lamanna, and Capitelli}}]{Celiberto1994}
\bibinfo{author}{\bibfnamefont{R.}~\bibnamefont{Celiberto}},
  \bibinfo{author}{\bibfnamefont{U.~T.} \bibnamefont{Lamanna}},
  \bibnamefont{and}
  \bibinfo{author}{\bibfnamefont{M.}~\bibnamefont{Capitelli}},
  \bibinfo{journal}{Phys. Rev. A} \textbf{\bibinfo{volume}{50}},
  \bibinfo{pages}{4778} (\bibinfo{year}{1994}), ISSN \bibinfo{issn}{10502947}.

\bibitem[{\citenamefont{Zammit et~al.}(2017{\natexlab{b}})\citenamefont{Zammit,
  Savage, Colgan, Fursa, Kilcrease, Bray, Fontes, Hakel, and
  Timmermans}}]{ZSCFKBFH717}
\bibinfo{author}{\bibfnamefont{M.~C.} \bibnamefont{Zammit}},
  \bibinfo{author}{\bibfnamefont{J.~S.} \bibnamefont{Savage}},
  \bibinfo{author}{\bibfnamefont{J.}~\bibnamefont{Colgan}},
  \bibinfo{author}{\bibfnamefont{D.~V.} \bibnamefont{Fursa}},
  \bibinfo{author}{\bibfnamefont{D.~P.} \bibnamefont{Kilcrease}},
  \bibinfo{author}{\bibfnamefont{I.}~\bibnamefont{Bray}},
  \bibinfo{author}{\bibfnamefont{C.~J.} \bibnamefont{Fontes}},
  \bibinfo{author}{\bibfnamefont{P.}~\bibnamefont{Hakel}}, \bibnamefont{and}
  \bibinfo{author}{\bibfnamefont{E.}~\bibnamefont{Timmermans}},
  \bibinfo{journal}{Astrophys. J.} \textbf{\bibinfo{volume}{851}},
  \bibinfo{pages}{64} (\bibinfo{year}{2017}{\natexlab{b}}).

\bibitem[{\citenamefont{Dalgarno and Stephens}(1970)}]{Dalgarno1970}
\bibinfo{author}{\bibfnamefont{A.}~\bibnamefont{Dalgarno}} \bibnamefont{and}
  \bibinfo{author}{\bibfnamefont{T.~L.} \bibnamefont{Stephens}},
  \bibinfo{journal}{Astrophys. J.} \textbf{\bibinfo{volume}{160}},
  \bibinfo{pages}{L107} (\bibinfo{year}{1970}).

\bibitem[{\citenamefont{Fantz et~al.}(2000)\citenamefont{Fantz, Schalk, and
  Behringer}}]{Fantz2000a}
\bibinfo{author}{\bibfnamefont{U.}~\bibnamefont{Fantz}},
  \bibinfo{author}{\bibfnamefont{B.}~\bibnamefont{Schalk}}, \bibnamefont{and}
  \bibinfo{author}{\bibfnamefont{K.}~\bibnamefont{Behringer}},
  \bibinfo{journal}{New J. Phys.} \textbf{\bibinfo{volume}{2}}
  (\bibinfo{year}{2000}).

\bibitem[{\citenamefont{Morrison and Sun}(1994)}]{Morrison1994}
\bibinfo{author}{\bibfnamefont{M.~A.} \bibnamefont{Morrison}} \bibnamefont{and}
  \bibinfo{author}{\bibfnamefont{W.}~\bibnamefont{Sun}}, in
  \emph{\bibinfo{booktitle}{Comput. Methods Electron-Molecule Collisions}},
  edited by \bibinfo{editor}{\bibfnamefont{W.}~\bibnamefont{Huo}}
  \bibnamefont{and} \bibinfo{editor}{\bibfnamefont{F.~A.}
  \bibnamefont{Gianturco}} (\bibinfo{year}{1994}).

\bibitem[{\citenamefont{Tennyson et~al.}(2017)\citenamefont{Tennyson, Rahimi,
  Hill, Tse, Vibhakar, Akello-Egwel, Brown, Dzarasova, Hamilton, Jaksch
  et~al.}}]{Tennyson2017}
\bibinfo{author}{\bibfnamefont{J.}~\bibnamefont{Tennyson}},
  \bibinfo{author}{\bibfnamefont{S.}~\bibnamefont{Rahimi}},
  \bibinfo{author}{\bibfnamefont{C.}~\bibnamefont{Hill}},
  \bibinfo{author}{\bibfnamefont{L.}~\bibnamefont{Tse}},
  \bibinfo{author}{\bibfnamefont{A.}~\bibnamefont{Vibhakar}},
  \bibinfo{author}{\bibfnamefont{D.}~\bibnamefont{Akello-Egwel}},
  \bibinfo{author}{\bibfnamefont{D.~B.} \bibnamefont{Brown}},
  \bibinfo{author}{\bibfnamefont{A.}~\bibnamefont{Dzarasova}},
  \bibinfo{author}{\bibfnamefont{J.~R.} \bibnamefont{Hamilton}},
  \bibinfo{author}{\bibfnamefont{D.}~\bibnamefont{Jaksch}},
  \bibnamefont{et~al.}, \bibinfo{journal}{Plasma Sources Sci. Technol.}
  \textbf{\bibinfo{volume}{26}} (\bibinfo{year}{2017}).

\bibitem[{\citenamefont{Gorfinkiel et~al.}(2002)\citenamefont{Gorfinkiel,
  Morgan, and Tennyson}}]{Gorfinkiel2002}
\bibinfo{author}{\bibfnamefont{J.~D.} \bibnamefont{Gorfinkiel}},
  \bibinfo{author}{\bibfnamefont{L.~A.} \bibnamefont{Morgan}},
  \bibnamefont{and} \bibinfo{author}{\bibfnamefont{J.}~\bibnamefont{Tennyson}},
  \bibinfo{journal}{J. Phys. B At. Mol. Opt. Phys.}
  \textbf{\bibinfo{volume}{35}}, \bibinfo{pages}{543} (\bibinfo{year}{2002}).

\bibitem[{\citenamefont{Tennyson and Trevisan}(2002)}]{Tennyson2002}
\bibinfo{author}{\bibfnamefont{J.}~\bibnamefont{Tennyson}} \bibnamefont{and}
  \bibinfo{author}{\bibfnamefont{C.~S.} \bibnamefont{Trevisan}},
  \bibinfo{journal}{Contrib. to Plasma Phys.} \textbf{\bibinfo{volume}{42}},
  \bibinfo{pages}{573} (\bibinfo{year}{2002}), ISSN \bibinfo{issn}{08631042}.

\bibitem[{\citenamefont{Burke and Tennyson}(2005)}]{Burke2005a}
\bibinfo{author}{\bibfnamefont{P.~G.} \bibnamefont{Burke}} \bibnamefont{and}
  \bibinfo{author}{\bibfnamefont{J.}~\bibnamefont{Tennyson}},
  \bibinfo{journal}{Mol. Phys.} \textbf{\bibinfo{volume}{103}},
  \bibinfo{pages}{2537} (\bibinfo{year}{2005}).

\bibitem[{\citenamefont{Chakrabarti and Tennyson}(2007)}]{Chakrabarti2007}
\bibinfo{author}{\bibfnamefont{K.}~\bibnamefont{Chakrabarti}} \bibnamefont{and}
  \bibinfo{author}{\bibfnamefont{J.}~\bibnamefont{Tennyson}},
  \bibinfo{journal}{J. Phys. B At. Mol. Opt. Phys.}
  \textbf{\bibinfo{volume}{40}}, \bibinfo{pages}{2135} (\bibinfo{year}{2007}).

\bibitem[{\citenamefont{Brigg et~al.}(2014)\citenamefont{Brigg, Tennyson, and
  Plummer}}]{BTP14}
\bibinfo{author}{\bibfnamefont{W.~J.} \bibnamefont{Brigg}},
  \bibinfo{author}{\bibfnamefont{J.}~\bibnamefont{Tennyson}}, \bibnamefont{and}
  \bibinfo{author}{\bibfnamefont{M.}~\bibnamefont{Plummer}},
  \bibinfo{journal}{J.$\sim$Phys.$\sim$B At. Mol. Opt. Phys.}
  \textbf{\bibinfo{volume}{47}}, \bibinfo{pages}{185203}
  (\bibinfo{year}{2014}).

\bibitem[{\citenamefont{Chakrabarti and Tennyson}(2009)}]{Chakrabarti2009}
\bibinfo{author}{\bibfnamefont{K.}~\bibnamefont{Chakrabarti}} \bibnamefont{and}
  \bibinfo{author}{\bibfnamefont{J.}~\bibnamefont{Tennyson}},
  \bibinfo{journal}{J. Phys. B At. Mol. Opt. Phys.}
  \textbf{\bibinfo{volume}{42}} (\bibinfo{year}{2009}).

\bibitem[{\citenamefont{Chakrabarti et~al.}(2017)\citenamefont{Chakrabarti,
  Dora, Ghosh, Choudhury, and Tennyson}}]{Chakrabarti2017}
\bibinfo{author}{\bibfnamefont{K.}~\bibnamefont{Chakrabarti}},
  \bibinfo{author}{\bibfnamefont{A.}~\bibnamefont{Dora}},
  \bibinfo{author}{\bibfnamefont{R.}~\bibnamefont{Ghosh}},
  \bibinfo{author}{\bibfnamefont{B.~S.} \bibnamefont{Choudhury}},
  \bibnamefont{and} \bibinfo{author}{\bibfnamefont{J.}~\bibnamefont{Tennyson}},
  \bibinfo{journal}{J. Phys. B At. Mol. Opt. Phys.}
  \textbf{\bibinfo{volume}{50}} (\bibinfo{year}{2017}).

\bibitem[{\citenamefont{Stibbe and Tennyson}(1998{\natexlab{b}})}]{Stibbe1998}
\bibinfo{author}{\bibfnamefont{D.~T.} \bibnamefont{Stibbe}} \bibnamefont{and}
  \bibinfo{author}{\bibfnamefont{J.}~\bibnamefont{Tennyson}},
  \bibinfo{journal}{J. Phys. B At. Mol. Opt. Phys.}
  \textbf{\bibinfo{volume}{31}}, \bibinfo{pages}{815}
  (\bibinfo{year}{1998}{\natexlab{b}}), ISSN \bibinfo{issn}{09534075}.

\bibitem[{\citenamefont{{El Ghazaly} et~al.}(2004)\citenamefont{{El Ghazaly},
  Jureta, Urbain, and Defrance}}]{AJUD04}
\bibinfo{author}{\bibfnamefont{M.~O.~A.} \bibnamefont{{El Ghazaly}}},
  \bibinfo{author}{\bibfnamefont{J.}~\bibnamefont{Jureta}},
  \bibinfo{author}{\bibfnamefont{X.}~\bibnamefont{Urbain}}, \bibnamefont{and}
  \bibinfo{author}{\bibfnamefont{P.}~\bibnamefont{Defrance}},
  \bibinfo{journal}{J. Phys. B At. Mol. Opt. Phys.}
  \textbf{\bibinfo{volume}{37}}, \bibinfo{pages}{2467} (\bibinfo{year}{2004}).

\bibitem[{\citenamefont{Andersen et~al.}(1997)\citenamefont{Andersen, Johnson,
  Kella, Pedersen, and Vejby-Christensen}}]{AJKPV97}
\bibinfo{author}{\bibfnamefont{L.~H.} \bibnamefont{Andersen}},
  \bibinfo{author}{\bibfnamefont{P.~J.} \bibnamefont{Johnson}},
  \bibinfo{author}{\bibfnamefont{D.}~\bibnamefont{Kella}},
  \bibinfo{author}{\bibfnamefont{H.~B.} \bibnamefont{Pedersen}},
  \bibnamefont{and}
  \bibinfo{author}{\bibfnamefont{L.}~\bibnamefont{Vejby-Christensen}},
  \bibinfo{journal}{Phys. Rev. A} \textbf{\bibinfo{volume}{55}},
  \bibinfo{pages}{2799} (\bibinfo{year}{1997}).

\bibitem[{\citenamefont{Heinemann et~al.}(2017)\citenamefont{Heinemann, Fantz,
  Kraus, Schiesko, Wimmer, W{\"{u}}nderlich, Bonomo, Fr{\"{o}}schle, Nocentini,
  and Riedl}}]{Heinemann2017}
\bibinfo{author}{\bibfnamefont{B.}~\bibnamefont{Heinemann}},
  \bibinfo{author}{\bibfnamefont{U.}~\bibnamefont{Fantz}},
  \bibinfo{author}{\bibfnamefont{W.}~\bibnamefont{Kraus}},
  \bibinfo{author}{\bibfnamefont{L.}~\bibnamefont{Schiesko}},
  \bibinfo{author}{\bibfnamefont{C.}~\bibnamefont{Wimmer}},
  \bibinfo{author}{\bibfnamefont{D.}~\bibnamefont{W{\"{u}}nderlich}},
  \bibinfo{author}{\bibfnamefont{F.}~\bibnamefont{Bonomo}},
  \bibinfo{author}{\bibfnamefont{M.}~\bibnamefont{Fr{\"{o}}schle}},
  \bibinfo{author}{\bibfnamefont{R.}~\bibnamefont{Nocentini}},
  \bibnamefont{and} \bibinfo{author}{\bibfnamefont{R.}~\bibnamefont{Riedl}},
  \bibinfo{journal}{New J. Phys.} \textbf{\bibinfo{volume}{19}}
  (\bibinfo{year}{2017}).

\bibitem[{\citenamefont{Lamara et~al.}(2006)\citenamefont{Lamara, Hugon, and
  Bougdira}}]{Lamara2006}
\bibinfo{author}{\bibfnamefont{T.}~\bibnamefont{Lamara}},
  \bibinfo{author}{\bibfnamefont{R.}~\bibnamefont{Hugon}}, \bibnamefont{and}
  \bibinfo{author}{\bibfnamefont{J.}~\bibnamefont{Bougdira}},
  \bibinfo{journal}{Plasma Sources Sci. Technol.}
  \textbf{\bibinfo{volume}{15}}, \bibinfo{pages}{526} (\bibinfo{year}{2006}).

\bibitem[{\citenamefont{Mizuochi et~al.}(2012)\citenamefont{Mizuochi, Tokuda,
  Ogura, and Yamasaki}}]{Mizuochi2012}
\bibinfo{author}{\bibfnamefont{N.}~\bibnamefont{Mizuochi}},
  \bibinfo{author}{\bibfnamefont{N.}~\bibnamefont{Tokuda}},
  \bibinfo{author}{\bibfnamefont{M.}~\bibnamefont{Ogura}}, \bibnamefont{and}
  \bibinfo{author}{\bibfnamefont{S.}~\bibnamefont{Yamasaki}},
  \bibinfo{journal}{Jpn. J. Appl. Phys.} \textbf{\bibinfo{volume}{51}}
  (\bibinfo{year}{2012}).

\bibitem[{\citenamefont{Rauner et~al.}(2017)\citenamefont{Rauner, Briefi, and
  Fantz}}]{Rauner2017a}
\bibinfo{author}{\bibfnamefont{D.}~\bibnamefont{Rauner}},
  \bibinfo{author}{\bibfnamefont{S.}~\bibnamefont{Briefi}}, \bibnamefont{and}
  \bibinfo{author}{\bibfnamefont{U.}~\bibnamefont{Fantz}},
  \bibinfo{journal}{Plasma Sources Sci. Technol.} \textbf{\bibinfo{volume}{26}}
  (\bibinfo{year}{2017}).

\bibitem[{\citenamefont{Fantz et~al.}(2013)\citenamefont{Fantz, Schiesko, and
  W{\"{u}}nderlich}}]{Fantz2013}
\bibinfo{author}{\bibfnamefont{U.}~\bibnamefont{Fantz}},
  \bibinfo{author}{\bibfnamefont{L.}~\bibnamefont{Schiesko}}, \bibnamefont{and}
  \bibinfo{author}{\bibfnamefont{D.}~\bibnamefont{W{\"{u}}nderlich}},
  \bibinfo{journal}{AIP Conf. Proc.} \textbf{\bibinfo{volume}{1515}},
  \bibinfo{pages}{187} (\bibinfo{year}{2013}).

\bibitem[{\citenamefont{W{\"{u}}nderlich
  et~al.}(2016)\citenamefont{W{\"{u}}nderlich, Kraus, Fr{\"{o}}schle, Riedl,
  Fantz, and Heinemann}}]{Wunderlich2016a}
\bibinfo{author}{\bibfnamefont{D.}~\bibnamefont{W{\"{u}}nderlich}},
  \bibinfo{author}{\bibfnamefont{W.}~\bibnamefont{Kraus}},
  \bibinfo{author}{\bibfnamefont{M.}~\bibnamefont{Fr{\"{o}}schle}},
  \bibinfo{author}{\bibfnamefont{R.}~\bibnamefont{Riedl}},
  \bibinfo{author}{\bibfnamefont{U.}~\bibnamefont{Fantz}}, \bibnamefont{and}
  \bibinfo{author}{\bibfnamefont{B.}~\bibnamefont{Heinemann}},
  \bibinfo{journal}{Plasma Phys. Control. Fusion} \textbf{\bibinfo{volume}{58}}
  (\bibinfo{year}{2016}).

\bibitem[{\citenamefont{Mizuochi et~al.}(2007)\citenamefont{Mizuochi, Isoya,
  Niitsuma, Sekiguchi, Watanabe, Kato, Makino, Okushi, and
  Yamasaki}}]{Mizuochi2007}
\bibinfo{author}{\bibfnamefont{N.}~\bibnamefont{Mizuochi}},
  \bibinfo{author}{\bibfnamefont{J.}~\bibnamefont{Isoya}},
  \bibinfo{author}{\bibfnamefont{J.}~\bibnamefont{Niitsuma}},
  \bibinfo{author}{\bibfnamefont{T.}~\bibnamefont{Sekiguchi}},
  \bibinfo{author}{\bibfnamefont{H.}~\bibnamefont{Watanabe}},
  \bibinfo{author}{\bibfnamefont{H.}~\bibnamefont{Kato}},
  \bibinfo{author}{\bibfnamefont{T.}~\bibnamefont{Makino}},
  \bibinfo{author}{\bibfnamefont{H.}~\bibnamefont{Okushi}}, \bibnamefont{and}
  \bibinfo{author}{\bibfnamefont{S.}~\bibnamefont{Yamasaki}},
  \bibinfo{journal}{J. Appl. Phys.} \textbf{\bibinfo{volume}{101}},
  \bibinfo{pages}{1} (\bibinfo{year}{2007}).

\bibitem[{\citenamefont{Stibbe and Tennyson}(1999)}]{ST99}
\bibinfo{author}{\bibfnamefont{D.~T.} \bibnamefont{Stibbe}} \bibnamefont{and}
  \bibinfo{author}{\bibfnamefont{J.}~\bibnamefont{Tennyson}},
  \bibinfo{journal}{Astrophys. J.} \textbf{\bibinfo{volume}{513}},
  \bibinfo{pages}{147} (\bibinfo{year}{1999}).

\end{thebibliography}

\end{document}